\documentclass[draft,onecolumn]{IEEEtran}
\usepackage{TIT_Expu_ArXiv}

\usepackage[utf8]{inputenc}

\usepackage{url}
\title{A Refinement of Expurgation}

\author{Giuseppe Cocco, Albert Guill\'en i F\`{a}bregas and Josep Font-Segura
		\thanks{Giuseppe Cocco is with the Department of Signal Theory and Communications, Universitat Polit\`ecnica de Catalunya, 08034, Barcelona, Spain (e-mail: giuseppe.cocco@upc.edu).

Albert Guill\'en i F\`{a}bregas is with the Department of Engineering, University
of Cambridge, CB2 1PZ Cambridge, U.K., and also with the Department
of Information and Communication Technologies, Universitat Pompeu Fabra,
08018 Barcelona, Spain (e-mail: guillen@ieee.org).

Josep Font-Segura is with the Department of Information and Communication Technologies, Universitat Pompeu Fabra, 08018, Barcelona, Spain (e-mail: josep.font@ieee.org).

 This work was supported in part by the Ramon y Cajal fellowship program (grant RYC2021-033908-I) funded by 
the Spanish Government through MCIN/AEI/10.13039/501100011033 and the European Union ``NextGenerationEU'' Recovery Plan, the European
Research Council under ERC Agreement 725411 and by the Spanish
Ministry of Economy and Competitiveness under Grant PID2020-116683GB-C22.
}
	}

\begin{document}
\maketitle

\begin{abstract}
\MR{We show that for a wide range of channels and code ensembles with pairwise-independent codewords, with probability tending to $1$ with the code length, expurgating an arbitrarily small fraction of codewords \AV{from a randomly selected code} results in a code attaining the expurgated exponent. }
\end{abstract}

%%%%%%%%%%%%%%%%%%%%%%%%%%%%%%%%
\section{Preliminaries}

We consider the problem of reliable communication of $M_n$ equiprobable messages over noisy channels described by a random transformation $\Wnvec$, where $\x\in \mathcal{X}^n$ and $\y\in\mathcal{Y}^n$ are the channel input and output sequences, and $\mathcal{X}$ and $\mathcal{Y}$ are the input and output alphabets, respectively. Each message $m\in\{1,\dotsc,M_n\}$, where $M_n=\lceil 2^{nR}\rceil$, $R$ being the code rate, is mapped onto an $n$-length codeword $\x_m$ sent over the channel. The code is defined as $\CnallMdet=\{\x_1,\dotsc,\x_{M_n}\}$.
We denote with $\PecnmallMdet$ the error probability when codeword $m\in\{1,\dotsc,M_n\}$ from code $\CnallMdet$ is transmitted; similarly $\PecnallMdet=\frac{1}{M_n}\sum_{m=1}^{M_n}\PecnmallMdet$ denotes the average error probability of the code. 
Let $\CnallM=\{\X_1, \dotsc, \X_{M_n}\}$ be a random code, i.e., a set of $M_n$ random codewords generated with probability $\PP[\CnallM = \CnallMdet] = \PP[\X_1=\x_1,\dotsc,\X_{M_n}=\x_{M_n}]$. We assume that codewords are generated in a pairwise independent manner, that is, for any two indices $m,k\in\{1,\ldots,M_n\}, m\neq k$, {it holds that} $\PP[\X_m=\x_m,\X_{k}=\x_k]=Q^n(\x_m)Q^n(\x_k)$\AV{,} where $Q^n(\x_m)=\PP[\X_m=\x_m]$ is a probability distribution defined over $\mathcal{X}^n$.
 
Let $\PecnmallM$ and $\PecnallM$ be the random variables denoting the error probability of the $m$-th codeword for  random code $\CnallM$ and the average error probability of the code, respectively.
We denote the $n$-length error exponents of such random variables by $\EmnallM = - \frac{1}{n}\log \PecnmallM$ and $\EnallM =  - \frac{1}{n}\log \PecnallM$, respectively.
For some ensembles and channels the ensemble-average of the code error probability $\EE\bigl[\PecnallM\bigr]$ is known to decay exponentially in $n$ \cite{feinstein1955}. A lower bound on the error exponent $-\frac{1}{n}\log\EE\bigl[\PecnallM\bigr]$ is given by Gallager's multi-letter random coding exponent $\Ern$ {in}~\cite[Eq.~(5.6.16)]{gallagerBook}. \MR{For the \ac{DMC}, this bound is known to coincide with the sphere-packing upper bound {on the reliability function} \cite{SHANNON_SP_1967,fano1961transmission} in the high rate region.}

In \cite[Sec. 5.7]{gallager1965simple} Gallager  showed that, for some channels and ensembles, there exists a code with strictly higher error exponent than $\Ern$ at low rates. In order to show this, Gallager considered a pairwise-independent ensemble with $M_n'=2M_n-1$ codewords. Using Markov's inequality he showed that
\begin{eqnarray}\label{eq:Gallag}
\PP\Bigl [ \Pecnmall \geq 2^{\frac{1}{s}}\EE[\Pecnmall^s]^{\frac{1}{s}} \Bigr ]\leq \frac{1}{2}
\end{eqnarray}
for any $s>0$.
He then introduced the indicator function
\begin{align}\label{eq:condit_Gall}
&\varphi_m\bigl(\CnallMdet\bigr)\notag\\&\ =\begin{cases}
1		\ \text{ if } \PecnmallMdet < 2^{\frac{1}{s}}\EE\bigl[\PecnmallM^s\bigr]^{\frac{1}{s}}\\
0		\ \text{ otherwise}
\end{cases}
\end{align}
and showed that, using \eqref{eq:Gallag} and \eqref{eq:condit_Gall},  the following {inequality} holds
\begin{eqnarray}\label{eq:gall_sum}
\EE\left[\sum_{m=1}^{M_n'}\varphi_m(\Cnall)\right]
\geq  M_n.
\end{eqnarray}
From \eqref{eq:gall_sum} it follows that, since the average number of codewords that have a probability of error smaller than $2^{\frac{1}{s}}\EE\bigl[\Pecnmalldet^s\bigr]^{\frac{1}{s}}$  in a randomly generated code with $M_n'=2M_n-1$ codewords is at least $M_n$, there must exist a code having at least $M_n$ codewords, out of the $M_n'$, fulfilling this property. Thus, by removing (expurgating) the worst half of the codewords from the code with $M_n'$ codewords we obtain a new code with $M_n$ codewords, each of which satisfies the condition in the first line of the right-hand side in~\eqref{eq:condit_Gall}. Finally, restricting $s$ to $0<s\leq 1$, Gallager derives a lower bound on the exponent of $2^{\frac{1}{s}}\EE[\Pecnmalldet^s]^{\frac{1}{s}}$, given by
\begin{align}\label{eqn:expu}
	\Eexn = E_{\rm x}^n(\hat\rho_n,Q^n) - \hat\rho_n R,
\end{align}
where
\begin{align}\label{eqn:ex}
\Exnvarrho& =  -\frac{1}{n}\log \biggl( \sum_{\x}\sum_{\x'} Q^n(\x)Q^n(\x') Z_n(\x,\x')^\frac{1}{\rho}\biggr)^\rho,
\end{align}
$Z_n(\x,\x')=\sum_{\y}\sqrt{\Wnvec\Wnvecp}$ is the Bhattacharyya coefficient between codewords $\x,\x'\in\Xc^n$ while
\begin{align}
\hat\rho_n = \argmax_{\rho\geq 1} \bigl\{ E_{\rm x}^n(\rho,Q^n) - \rho R\bigr\}
\label{eq:rhohatn}
\end{align}
is the parameter that yields the highest exponent. The preceding argument is valid for the maximal probability of error, since every codeword in the expurgated code attains the same exponent. In addition, observe that \MR{since} \eqref{eq:gall_sum} uses the standard ensemble-average argument (i.e.~by taking the average over the ensemble) we show the existence of a code with the desired property.
The exponent in~\eqref{eqn:expu} is the expurgated exponent. We refer to the code with $M_n'$ codewords before expurgation as a mother code. We say that a mother code is good if, once expurgated, we obtain a code with asymptotically the same rate, the codewords of which each have an exponent at least as large as the expurgated.

A refinement of the above follows from \eqref{eq:Gallag}. Specifically, for $\epsilon>0$ it can be shown that there exists a code with $M_n'=M_n(1+\epsilon)$ codewords such that removing $\epsilon M_n$ codewords  yields a code that attains the expurgated exponent \cite[Lemma 1]{scarlett_TIT2014}. Although \cite[Lemma 1]{scarlett_TIT2014} \MR{generalizes} Gallager's method, \MR{it still only shows} the existence of a code that attains the expurgated exponent.

\section{Main Result}\label{sec:main}
This paper strengthens existing results on expurgation by showing that the probability of finding a code with $M_n'=(1+\epsilon)M_n$ codewords that contains a code with at least $M_n$ codewords each of which achieving the expurgated exponent tends to $1$ with the code length. \MR{We define the sequence  $\delta_n = \frac{\hat\rho_n}{n} \log \gamma_n$, where $\gamn$ is such that $\lim_{n\rightarrow\infty} \gamn=\infty$ while $\lim_{n\rightarrow\infty} \frac{\log\gamn}{n}=0$, $\hat\rho_n$ being a positive sequence defined in \eqref{eq:rhohatn} that depends on \AV{the channel, the ensemble and the rate}. From the definition of $\delta_n$ it can be seen that if $\hat{\rho}_n$ either converges to a constant or grows sufficiently slowly, there exists a $\gamn$ such that $\delta_n\rightarrow 0$.} Similarly to Gallager, for a given $\delta_n$, we define the  indicator function
\begin{align}\label{eq:condit}
\phi_m\bigl(\CnallMdet\bigr)=\begin{cases}
1		\ \text{ if } \EmnallMdet > \Eexn -\delta_n\\
0		\ \text{ otherwise},
\end{cases}
\end{align}
and the number of codewords attaining an exponent higher than $\Eexn -\delta_n$ as
\begin{eqnarray}\label{eq:numcodws}
\Phi\bigl(\Cnalldet\bigr)~{=}\sum_{m=1}^{M_n'}\phi_m\bigl(\Cnalldet\bigr).
\end{eqnarray}

\begin{theorem}\label{theo:1}
Consider a pairwise-independent code \MR{ensemble} with $M_n'=M_n(1+\epsilon)$ codewords and any $\epsilon>0$. If the  sequence $\{\delta_n\}_{n=1}^\infty$, which depends on the channel and the ensemble, satisfies $\lim_{n\rightarrow\infty}\delta_n=0$, then for any $0<\epsilon_1<\epsilon$, it holds that
 \begin{align}\label{eq:statement}
\lim_{n\rightarrow\infty}\PP\bigl[\Phi\bigl(\Cnall\bigr)\geq M_n(1+\epsilon_1)\bigr]=1.
\end{align}

\end{theorem}

\begin{IEEEproof}
See Section \ref{sec:proof}.
\end{IEEEproof}

In words, with high probability we find a mother code with $M_n'=(1+\epsilon)M_n$ codewords, $M_n$ of which attain the expurgated exponent. That is, good mother codes are found easily and only contain an arbitrarily small fraction $\epsilon/(1+\epsilon)$ of codewords that need to be expurgated. Theorem \ref{theo:1} extends Gallager's method, and applies\MR{, among others,} to \ac{i.i.d.} and constant composition codes over \ac{DMC}s, as well as  channels with memory such as the finite-state channel in \cite[Sec.~4.6]{gallagerBook}, for which the expurgated exponent is derived in \cite{coccoTIT2022}. 

As a final remark, recent  works \cite{merhav_TIT2018,LargeDevLogErrProb_tamir_TIT2020,coccoTIT2022,TruongGJF2022a} show that for many ensembles, most low-rate codes have an error exponent  $\EnallM$ that is strictly larger than the exponent of the ensemble average error probability, i.e., the random coding exponent. Similarly, Theorem \ref{theo:1} implies that for most codes, almost any codeword has an associated error exponent $\EmnallM$ that is strictly larger than the ensemble average of the exponent of the error probability of the codebook $\EE\big[\EnallM \big]$. In both cases the smaller error exponent of the average probability of error is due to a relatively small number of elements (codes in the first case, codewords in the second) that perform poorly. 
\MR{Furthermore, as shown in \cite{LargeDevLogErrProb_tamir_TIT2020,TruongGJF2022a} for \ac{i.i.d.} and constant composition codes over \ac{DMC},} the error exponents of the codes in the ensemble concentrate around the \ac{TRC} exponent \cite{barg_forney_TIT2002,merhav_TIT2018}. Similarly to such works, it can be shown that the error exponent $\EmnallM$, for any $m$, concentrates around its mean, the expurgated exponent. The proof makes use of Lemma \ref{th:lb_exp} in Section \ref{sec:proof}, and follows almost identical steps as in \cite[Theorem 1]{TruongGJF2022a}, \cite[Theorem 1]{coccoTIT2022} and \cite[Theorem 2]{coccoTIT2022} once $\Pecns$ is replaced by $\Pecnmdet$ and it is omitted here.

%%%%%%%%%%%%%%%%%%%%%%%%%%%%%%%%%%
\section{Proof of Theorem 1}\label{sec:proof}
We start with the following lemma, whose proof is almost identical to that of \cite[Lemma 1]{coccoTIT2022}.
\begin{lemma}\label{th:lb_exp}
For a channel $W^n$ and a pairwise-independent $M_n'$-codewords code ensemble with codeword distribution $Q^n$, for any ${m}\in \{1,\ldots,M_n'\}$ it holds that
\begin{equation}\label{eqn:theo5_statement}
\mathbb{P}\bigl[ \Emnall > \Eexn  - \delta_n \bigr]\geq 1-\frac{1}{\gamn},
\end{equation}
where $\gamma_n$ and $\delta_n$ are positive real-valued sequences. 
\end{lemma}

The proof of Lemma \ref{th:lb_exp} follows from Markov's inequality
\begin{eqnarray}\label{eqn:lemma1}
\PP\Bigl [ \Pmcn \geq \gamn^{\frac{1}{s}}\EE[\Pmcn^s]^{\frac{1}{s}} \Bigr ]\leq \frac{1}{\gamn}
\end{eqnarray}
and applying the same steps as in \cite[Theorem 1]{coccoTIT2022} once $\Pecn$ is replaced with $\Pmcn$. \MR{The sequences $\gamma_n$ and $\delta_n$ are the same as those introduced in Section \ref{sec:main}}. 
\MR{Observe that using inequality \eqref{eqn:lemma1} and following similar steps as in \cite{coccoTIT2022} it can be shown that $\lim_{n\rightarrow\infty}\Eexn$ is a lower bound on $\lim_{n\rightarrow\infty}\EE[\EmnallM]$. Furthermore, using similar arguments as in \cite{TruongGJF2022a} it can be shown that such bound is tight at least for \ac{i.i.d.} and constant composition codes over \ac{DMC}. That is, for such ensembles and channels  $\lim_{n\rightarrow\infty} \EE\big[-\frac{1}{n}\log \PecnmallM\big] = \lim_{n\rightarrow\infty}\Eexn$, i.e., the expurgated is the typical codeword exponent.}

\MR{If the positive sequence $\hat\rho_n$, defined in \eqref{eq:rhohatn}, converges or grows sufficiently slowly, then there exists a sequence $\gamn$ such that $\lim_{n\rightarrow\infty} \gamn=\infty$,  $\lim_{n\rightarrow\infty} \frac{\log\gamn}{n}=0$, for which $\delta_n = \frac{\hat\rho_n}{n} \log \gamma_n\rightarrow 0$. \AV{For rate zero}, that is when $\lim_{n\to\infty} \frac{1}{n}\log{M_n}=0$, the $n$-length error exponent in~\eqref{eqn:expu} depends on the particular subexponential growth of $M_n$, while $\hat\rho_n$ tends to infinity \AV{with} a growth that depends on the channel and the ensemble. In this case, as discussed in the paragraph succeeding \cite[Eq.~(89)]{coccoTIT2022}, the assumption that $\frac{\hat\rho_n}{n} \log \gamma_n\rightarrow 0$ holds if the normalized variance of the Bhattacharyya coefficient $Z_n(\x,\x')$ grows slower than $\smash{\sqrt{\frac{n}{\log\gamma_n}}}$. In any case, choosing such $\gamma_n$ and applying Lemma \ref{th:lb_exp} we have that}
\begin{equation}\label{eqn:expu_common1}
\mathbb{P}\bigl[ \Emnall > \Eexn -\delta_n \bigr]\geq 1-\frac{1}{\gamn}.
\end{equation}
The random variable $\Phi(\Cnall)$, averaged across the ensemble, satisfies
\begin{align}\label{eq:ineq_sum}
\EE[\Phi(\Cnall)]&=\sum_{m=1}^{M_n(1+\epsilon)}\EE[\phi_m(\Cnall)]\\\label{eq:ineq_sum1}
&\geq \sum_{m=1}^{M_n(1+\epsilon)}\left(1-\frac{1}{\gamn}\right)\\
&=M_n(1+\epsilon)\left(1-\frac{1}{\gamn}\right),\label{eq:ineq_sum3}
\end{align}
where \eqref{eq:ineq_sum1} follows from the definition of the indicator function \eqref{eq:condit} and \eqref{eqn:expu_common1}.

\MR{We define $\Psi(\Cnall) = M_n' - \Phi(\Cnall)$, which is the number of codewords with exponent smaller than $\Eexn -\delta_n$. From \eqref{eq:ineq_sum3} it follows that
\begin{align}\label{eq:ineq_sum_bad}
\EE[\Psi(\Cnall)]\leq\frac{M_n(1+\epsilon)}{\gamn}.
\end{align}
Then, for sufficiently large $n$ we have that
\begin{align}
\PP\Big[\Psi(\Cnall)>\frac{M_n(1+\epsilon)}{\sqrt{\gamn}}\Big]
&~\AV{\leq}~\frac{1}{\sqrt{\gamn}},\label{eq:bad_markov_2}
\end{align}
where \eqref{eq:bad_markov_2} follows from Markov's inequality and \eqref{eq:ineq_sum_bad}. 
This shows that the probability of finding a code with many codewords with exponent strictly smaller than $\Eexn -\delta_n$ vanishes with $n$.} \AV{To prove our main result, we write the tail probability in~\eqref{eq:statement} as 
\begin{align}\label{eq:chain_good1}
&\PP\Big[\Phi(\Cnall)\geq M_n(1+\epsilon_1)\Big]\notag\\&\ = 1-\PP\Big[\Phi(\Cnall) < M_n(1+\epsilon_1)\Big]\\
&\ = 1-\PP\Big[\Psi(\Cnall)>M_n(\epsilon-\epsilon_1)\Big],
\end{align}
where we used the definitions of $\Psi(\Cnall)$ and $M_n'$. Since $\gamma_n$ tends to infinity, there must exist an $n_0\in\mathbb{N}$ such that $\epsilon-\epsilon_1>\frac{(1+\epsilon)}{\sqrt{\gamma_n}}$ for $n>n_0$ and therefore
\begin{align}
&\lim_{n\to\infty}\PP\Big[\Phi(\Cnall)\geq M_n(1+\epsilon_1)\Big]\notag\\ 
&\ \geq \lim_{n\to\infty} 1-\PP\Big[\Psi(\Cnall)>\frac{M_n(1+\epsilon)}{\sqrt{\gamma_n}}\Big]\label{eq:chain_good1_a}\\
&\ \geq  \lim_{n\to\infty} 1-\frac{1}{\sqrt{\gamn}}, \label{eq:chain_good1_b}
\end{align}
where \eqref{eq:chain_good1_b} follows from \eqref{eq:bad_markov_2}.
Finally, solving the limit yields the desired result.}

\bibliographystyle{IEEEtran}
\bibliography{IEEEabrv,TIT_Expu_ArXiv}

\end{document}